\documentclass[%
 reprint,
 amsmath,amssymb,
 aps,
]{revtex4-1}

\usepackage{graphicx}
\usepackage{dcolumn}
\usepackage{bm}
\usepackage{amsmath}
\usepackage{mathtools}
\usepackage{soul}
\usepackage{color}
\newcommand{\norm}[1]{\left\lVert#1\right\rVert}

\makeatletter
\def\thickhline{%
  \noalign{\ifnum0=`}\fi\hrule \@height \thickarrayrulewidth \futurelet
   \reserved@a\@xthickhline}
\def\@xthickhline{\ifx\reserved@a\thickhline
               \vskip\doublerulesep
               \vskip-\thickarrayrulewidth
             \fi
      \ifnum0=`{\fi}}
\makeatother

\newlength{\thickarrayrulewidth}
\begin{document}

\title{On the (im)possibility of extending the GRW model to relativistic particles}

\author{C. Jones}
 \email{caitlinisobel.jones@phd.units.it}
 \affiliation{Department of Physics, University of Trieste,
Strada Costiera 11, 34151 Trieste, Italy
}%
 \affiliation{Istituto Nazionale di Fisica Nucleare, Trieste Section, Via Valerio 2, 34127 Trieste, Italy}
 
\author{T. Guaita}%
 \email{tommaso.guaita@mpq.mpg.de}
\affiliation{Max Planck Institute of Quantum Optics, Hans-Kopfermann-Str. 1, D-85748 Garching, Germany\
}%

\author{A. Bassi}
 \email{bassi@ts.infn.it}
 \affiliation{Department of Physics, University of Trieste,
Strada Costiera 11, 34151 Trieste, Italy\
}%
 \affiliation{Istituto Nazionale di Fisica Nucleare, Trieste Section,  Via Valerio 2, 34127 Trieste, Italy}

\date{\today}

\begin{abstract}
		Spontaneous collapse models are proposed modifications to quantum mechanics which aim to solve the measurement problem. 
In this article we will consider models which attempt to extend a specific spontaneous collapse model, the  Ghirardi-Rimini-Weber model (GRW), to be consistent with special relativity. We will present a condition that a relativistic GRW model must meet for three cases: for single particle, for N distinguishable particles, and for indistinguishable particles. We will then show that this relativistic condition implies that one can have a relativistic GRW model for a single particles or for distinguishable non-interacting, non-entangled particles but not otherwise.
\end{abstract}

\maketitle

	\section{Introduction} \label{Introduction}
In quantum mechanics there are two forms of dynamics; unitary evolution, which is time reversible and preserves superpositions, which describes the evolution of isolated systems, and evolution described by positive operator valued measures (POVMs) which describes a system undergoing a measurement. The measurement problem is the fact that quantum mechanics fails to provide a precise description of which form of evolution describes any one situation. From observation limits can be placed on which regimes may be described with unitary evolution or POVMs, but the theory itself does not provide these. \\
Quantum field theory, the version of quantum mechanics consistent with special relativity, suffers from the same conceptual issue as non-relativistic quantum mechanics. It is a mathematical tool for calculating the probability of an outcome of a measurement given a initial condition, but it does not have prescription for when to use unitary or non-unitary dynamics. 

Since its discovery there have been many attempts to solve the measurement problem, some of the most famous suggestions include the many worlds interpretation \cite{vaidman2002many, everett1957relative} and Bohmian mechanics \cite{bohm1952suggested, PhysRev.85.180,teufel2009bohmian}. Both of these suggestions are experimentally indistinguishable from standard quantum mechanics.  \\
Spontaneous collapse models, first introduced by Ghirardi-Rimini-Weber  \cite{ghirardi1986unified} and Pearle \cite{pearle1976reduction}, solve the measurement problem by giving a new dynamics which completely describes the time evolution of the system at a non-relativistic level. They offer different experimental predictions than standard QM, and there are currently experiments investigating these effects \cite{piscicchia2017csl, vinante2017improved, carlesso2016experimental,zheng2019room}. The new dynamics is defined by introducing additional stochastic non-linear terms to the  Schr{\"o}dinger equation. These terms alter the form of unitary evolution such that there is a non-zero probability of the wavefunction describing a particle undergoing a spontaneous spatial localisation. This rate is proposed to be extremely low, such that a single particle  may remain in a superposition for a long period of time, in line with what is seen experimentally. However for multiple particles which are entangled then any single particle spontaneously collapsing collapses all particles it is entangled with. This effectively increases the rate of collapse for systems with high numbers of particles, such that macroscopic bodies are localised on extremely short time scales. This is often called the amplification mechanism and it ensures macroscopic classicality. This removes the need for the theory to include a description of an external observer, as macroscopic measuring apparatus interacting with a microscopic system causes the microscopic system to become entangled and hence collapse, via the amplification mechanism. For a full review of spontaneous collapse models see \cite{bassi2003dynamical}.  \\
In order for a spontaneous collapse model to be a successful description of the underlying physics it must be consistent with special relativity.  There is a tension between quantum mechanics and special relativity as quantum mechanics is non-local because space-like separated measurements of entangled systems must be correlated (as argued by EPR in \cite{einstein1935can}). A spontaneous collapse model should predict non-local correlations in order to remain consistent with experiment.

This paper is concerned with collapse model's consistency only with special relativity.
From now on in this article we will use relativistic to mean consistent with special relativity.  
\\
In its original formulation the GRW model was not relativistic and described distinguishable particles with discrete points of localisation. Continuous time collapse models have also been developed for instance in \cite{csl,tumulka2006spontaneous}.  There are various proposed models for relativistic collapse models: \cite{bedingham2014matter} where a prescription for the probability distribution of a matter density operator is Lorentz invariant, \cite{bedingham2011relativistic} which introduces a  mediating pointer field, \cite{tilloy2017interacting} in which collapse dynamics emerge by tracing out an environment from a relativistic quantum field theory, \cite{nicrosini2003relativistic} which proposes that the terms modifying the conventional Schr{\"o}dinger equation are functions of the stress-energy tensor. Pearle suggested a model \cite{pearle1999relativistic} and a proposed alteration of this in \cite{oppenheim2009fundamental} where energy is conserved by considering relational collapses. In \cite{dowker2004spontaneous} a collapse model on a $1+1$ lattice is presented and the authors suggest that it may be relativistic in the continuum limit. \\

 In this paper we will ask if it is possible for GRW to be made consistent with relativity whilst retaining its characteristic features.
 
 For single particles, distinguishable and indistinguishable particles  we will apply the conditions for consistency with relativity to the case of GRW and consider if such conditions permit models which give rise to collapses which localise the wavefunction and cause classical behaviour to emerge at large scales. We discuss where an existing model fits into this framework \cite{tumulka2006relativistic}. Something that has been under analysed in the relativistic collapse model literature is the fact that in order to be consistent with special relativity is it not sufficient to only ask that the dynamics are Lorentz covariant, it is also required that initial conditions between two inertial frames can  be compared. In this article we pay special attention to this fact and show how this limits the possible extensions of GRW.

This paper is organised as follows: in section \ref{qm_and_sr} the relationship between special relativity and quantum mechanics is reviewed and the Tomogana-Schwinger formalism is discussed. In section \ref{orgin_GRW} the original GRW model is introduced in the  Tomogana-Schwinger formalism. In section \ref{collapse_sr} relativistic conditions for single particle distinguishable and indistinguishable particle GRW models are given. For the indistinguishable  case it is shown that a such a model is either not relativistic or does not achieve macroscopic classicality.

\section{Quantum Mechanics and Special Relativity} \label{qm_and_sr}

Standard quantum mechanics (ignoring the measurement problem) provides probability distributions for the values of observables that are measured. A relativistic quantum mechanics must predict that observers in any two inertial frames to have the same measurement statistics for the outcome of any experiment they can perform. This is the conclusion reached in \cite{aharonov1980states} by Aharonov and Albert. They state that for a system with observables  $A, B, C...$  each with potential values $a, b, c...$  where observable $A$ is measured at time $t_a$ and found to have the value $a$, and other variables respectively, then agreement with special relativity implies that there is a covariant way of calculating the probability $P$ of $P_{t_a, t_b}(a, b,... |c, t_c, d, t_d... )$ i.e.  

\begin{multline}
\label{covar_prob}
P_{t_a, t_b}(a, b,... |c, t_c, d, t_d... )
 =\\ P_{t'_a, t'_b}(a', b',... |c', t'_c, d', t'_d... ).
\end{multline}

where $a'$, $b'$ etc. are the values of the observables in  the coordinates of a different inertial frame.
 This condition is \textit{stronger} than only requiring that the equations of motion transform covariantly, as in order to check this condition {\it one must be able to compare the initial conditions in each frame} \footnote{This requirement for consistency with SR in the active transformation rather that the passive viewpoint, because to us it seems the most natural viewpoint when considering stochastic dynamics}.  This requirement is inline with the usual definition of Lorentz covariance for quantum field theory \cite[Chapter 3]{greiner1990relativistic},  where is stated that for Lorentz covariance of a theory there must be an explicit rule for one observer to find their state of the system given the state of the same system in a different inertial frame (and of course the dynamics must be Lorentz invariant).
 
We choose to use this definition instead of other ways of characterising `relativistic' as it ensures that observers in any two frames are guaranteed to obtain the same result of an experimental run.

\begin{figure}
  	   		\centering
  	   		\includegraphics[width=1\columnwidth]{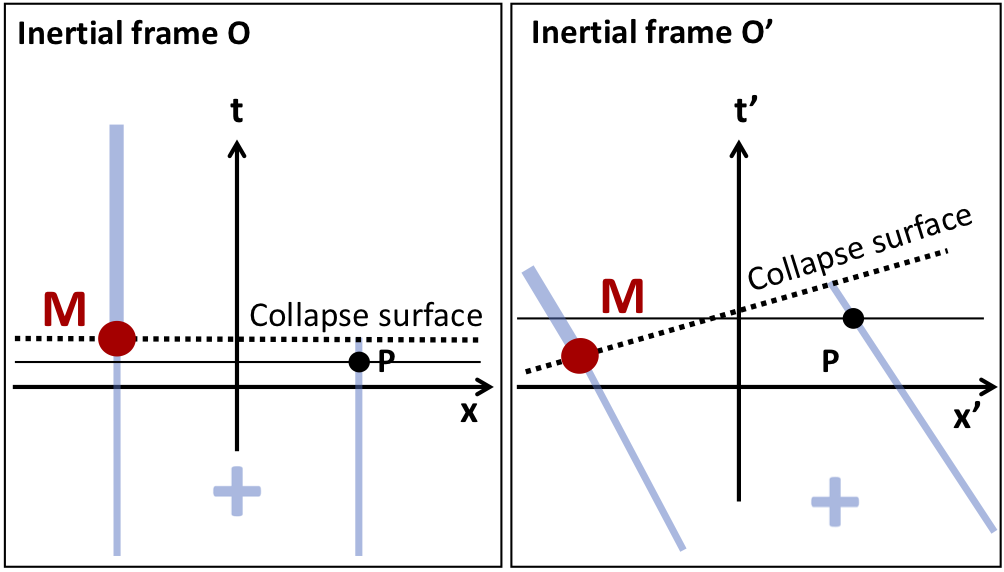}
  	   		\caption{A spacetime diagram showing the support of the wavefunction before and after a measurement $M$ where the wavefunction is a function over all of spacetime. The support is the shaded line with the amplitude proportional to the thickness of the line. The point $P$ is a  spacetime point of interest.  Suppose in one frame (left figure) the wavefunction is initially in a spatial superposition (as seen in that the support is present in two places and as denoted with the plus), then $M$ occurs and the wavefunction collapses along a specific constant time hypersurface (dotted line). The wavefunction on the surface intersecting point $P$ (thin black line) is normalised. However in a different inertial frame (right figure) if the collapse occurs along the {\it same} hypersurface (dotted line), then  the wavefunction on the constant time hyperplane intersecting point $P$ in the new frame (thin black line) is not normalised. } 
  	   		\label{not_normalised}
  	   	\end{figure}

 Probability distributions in quantum mechanics are found from the wavefunction via the Born rule. For non-relativistic quantum mechanics single particle wavefunctions are functions over every point in spacetime. However if one wishes to have a relativistic quantum mechanics where the wavefunction undergoes instantaneous collapses triggered by measurements, then this is not possible, as this implies that the wavefunction will not be normalised on a constant time hyperplane in some inertial frames, see Figure \ref{not_normalised}. Also, a preferred frame is selected, the one where the collapse occurs instantaneously. On the other hand, instantaneous collapses are required to ensure that non-local observables (for example momentum or total charge) are conserved \cite{aharonov1980states, aharonov1984usual}. Additionally instantaneous collapse of the state vector is required to ensure that Bell's inequalities are  violated. This is because in order for the outcomes of the Bell experiment to be perfectly correlated even though the results are from  space-like measurements, the two wings of the experiment must effect each other instantly, hence the wavefunction must collapse instantly \footnote{Hence proposals like \cite{hellwig1970formal} where collapse only effects the future light cone fail as they do not predict non-local correlations between outcomes of experiments. Since these non-local correlations are observed in nature then any successful theory must predict them.}.

  	   	In order to offer a frame independent description of instantaneous collapse of the wavefunction, Aharonov and Albert proposed an alternative way of describing the collapse when a measurement is performed~\cite{aharonov1981can}, in which the wavefunction collapses instantaneously in \textit{every} inertial frame.  
To allow the wavefunction to collapse instantaneously in every frame, it must be defined not on the 4D manifold but on  space-like 3D hypersurfaces which make up the manifold. Then the wavefunction, and hence normalised state in a Hilbert space can be defined on each hypersurface. 

\begin{figure}
   	   		\centering
   	   		\includegraphics[width=1\columnwidth]{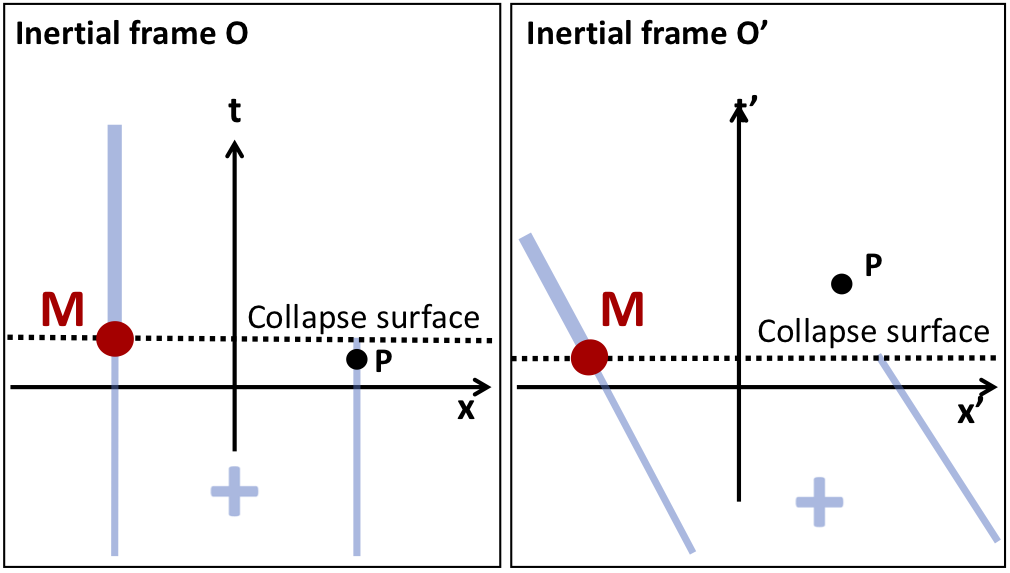}
   	   		\caption{A spacetime diagram showing the support of the wavefunction before and after a measurement $M$. Here, in every inertial frame the wavefunction collapses on a constant time hypersurface (dotted line) so that the wavefunction is always normalised for all observers. Note that the amplitude of the wavefunction at $P$ in different frames differs, this is a consequence of treating the wavefunction as a function on a 3D hypersurface. } 
   	   		\label{normalised}
   	   	\end{figure}

 Wavefunctions are defined on space-like hypersurfaces, if we label a hypersurface as $\omega$ then we can write a wavefunction on it as $ \psi_\omega(x) $. The coordinate $x$ here labels the coordinates of the 3D surface $\omega$  but is a four vector  $x \in  \mathbb{M}^4$ as $\omega$ is understood to be embedded in 4D spacetime. So then every inertial observer has a wavefunction defined on their constant time 3D hypersurface. However each state may have different values at the same specific space-time point $X$  so that $\psi_{\omega}(X)  \neq \psi_{\omega'}(X')$ where $X$ and $X'$ are the same point in two different inertial frames. This allows wavefunctions to be normalised in every frame, see Figure \ref{normalised}. This is acceptable because the wavefunction in this framework has no ontological meaning, it is simply a tool for calculating probabilities for the value of observables.

 In this framework every inertial observer can describe the time evolution of their system in terms of wavefunctions on parallel constant time hypersurfaces within their frame using the Tomogana-Schwinger formalism. We will introduce this formalism and show that if collapses are excluded, then this description is Lorentz covariant if it is integrable. Then for the case of quantum mechanics with measurements we will derive a condition on the measurement operator for Lorentz covariance in this framework.  
   	   	
 \subsection{The Tomogana-Schwinger formalism}
The Tomogana-Schwinger formalism \cite{tomonaga1946relativistically,schwinger1948quantum} describes unitary evolution as maps between wavefunctions defined on arbitrary space-like hypersurfaces without collapses. First we will introduce some additional notation for hypersurfaces. Let $\omega$ signify any generic space-like 3 dimensional hypersurface, let $\sigma_t$  denote a constant time hyperplane at time $t$ in a inertial frame $\mathcal{F}$ and hence $\sigma'_{t'}$ is a constant time hyperplane in a different inertial frame $\mathcal{F'}$.
 Then suppose the wavefunction is defined on an $\omega$ in the manifold $\mathbb{M}^4$. In this article we restrict ourselves to considering Minkowski spacetime $\mathbb{M}^4$ as it is sufficient to see the relevant Lorentz transformation properties of the probability distributions. Then in inertial frame $\mathcal{F}$  which has coordinates $x$  on a hypersurface $\omega$ the wavefunction is $ \psi_\omega(x)$.  In another inertial frame $\mathcal{F'}$ with coordinates $x'$ then on the same hyperplane $\omega$ the wavefunction is written $ \psi'_\omega(x')  $. A wavefunction under a Lorentz boost transforms as:
 \begin{equation} \label{state_transform}
  \psi_\omega(x) \to	 \psi'_\omega(x')  =   \Lambda   \psi_\omega(x) .
 \end{equation}
where $\Lambda$ is a representation of the Lorentz group. In other words on the same hypersurface the wavefunctions are equivalent up to a Lorentz transform.

\begin{figure}
   	   		\centering
   	   		\includegraphics[width=0.7\columnwidth]{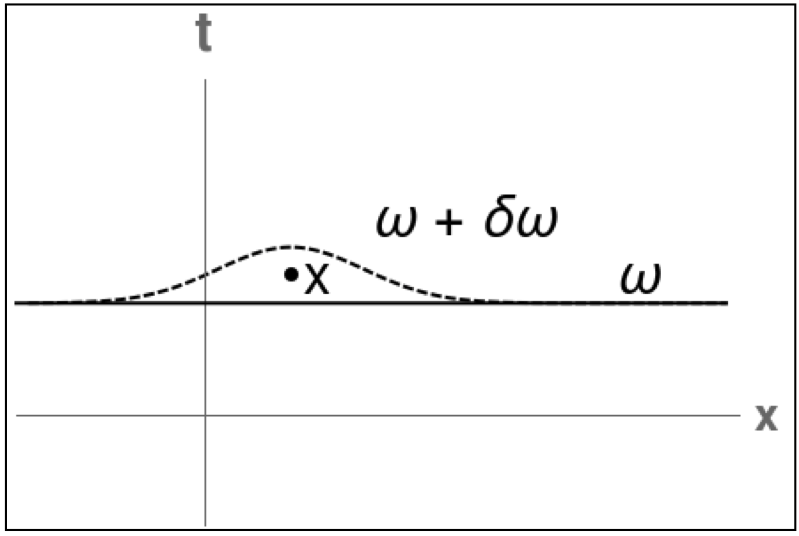}
   	   		\caption{A diagram showing the infinitesimal variation, $\delta \omega$, of the hypersurface $\omega$ about the point $x$} 
   	   		\label{func_deriv}
   	   	\end{figure}

 Analogously to the Schr{\"o}dinger equation, Tomogana and Schwinger defined the evolution of a wavefunction as it evolves between hypersurfaces, if there are no measurements between those surfaces:

 \begin{equation}
 \label{ts_eq}
 \frac{\delta}{\delta \omega(x)}\psi_\omega(x)  = -i \mathcal{H}(x) \psi_\omega(x) 
 \end{equation}
 where $ \frac{\delta}{\delta \omega(x)}$ is the functional derivative with respect to $\omega$ and $\mathcal{H}(x)$ is the Hamiltonian density. The functional derivative can be understood to be the variation in $\psi_\omega(x)$ with respect to a infinitesimal variation of $\omega$ about point $x$, see figure \ref{func_deriv}. 
The integrability condition for this system is that $[\mathcal{H}(x), \mathcal{H}(y)]=0$ if $x$ and $y$ are space-like separated. Equation \eqref{ts_eq} gives rise to an unitary evolution operator which relates two hypersurfaces:
    \begin{equation} 
    \label{time_ordered_unitary}
        U_{\omega_1}^{\omega_2} = T  \:\exp\left[-i \int_{\omega_1}^{\omega_2} d^4x \mathcal{H}(x)\right]
        \end{equation}
    such that  $ \psi_{\omega_2} (x) = U_{\omega_1}^{\omega_2} \psi_{\omega_1} (x)$, where $T$ means time ordering with respect to the frame  $\mathcal{F}$. This operator is frame independent although $\mathcal{H}(x)$ is not Lorentz invariant, the only frame dependant terms from the time ordering are zero due to the integrability condition \cite{ breuer1998relativistic, koba1950integrability}.   Therefore we have that for a frame $\mathcal{F}'$:
          \begin{subequations}
          	\begin{align} 
          	U_{\omega_1}^{' \omega_2} =& T'  \:\exp\left[-i \int_{\omega_1}^{\omega_2} d^4x' \mathcal{H'}(x')\right]  \label{time_ordered_unitary_primed} \\
          	=&\Lambda^\dagger U_{\omega_1}^{\omega_2}\Lambda \label{unit_lorentz}
          	\end{align}
          \end{subequations}

 \subsection{The Tomogana-Schwinger formalism with measurements} \label{ts_with_measurements}

  Now we wish to extend this formalism to describe collapses of the wavefunction due to measurements. In this article we will consider only collapses in the spatial basis as this is sufficient to explain the values of any experiment performed, as any observable can be coupled to position \cite{bassi2007quantum}.
  
  In a frame $\mathcal{F}$ the spatial collapse of the wavefunction at  $x \in \mathbb{M}^4$  is described though an operator $\hat{L}_{\omega}(x)$ defined on the Hilbert space on a space-like hypersurface $\omega$ passing through $x$. Following Albert and Aharanov we consider the  that the collapse occurs on the constant time hyperplane intersecting $x$, labelled $\sigma_t$ where $t=x_0$. This means that the collapse is described as occurring instantaneously in $\mathcal{F}$, as discussed in section \ref{qm_and_sr}. 
  
  $\hat{L}_{\sigma_t}(x)$ localises the particle it acts on about $x$ (if the wavefunction is not already localised). The properties of this operator are model dependant however in general it is not unitary. In a different frame $\mathcal{F}'$ the collapse operator $\hat{L'}_{\sigma'_{t'}}(x')$ is defined on a constant time hypersurface  $\sigma'_{t'}$. 
      
\begin{figure}
    	   	  	   		\centering
    	   	  	   		\includegraphics[width=0.7\columnwidth]{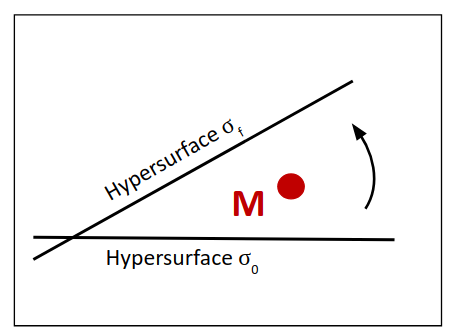}
    	   	  	   		\caption{A schematic showing a measurement at a point $M$ between two hypersurfaces $\sigma_0$ and $\sigma_f$. It is not possible to relate wavefunctions on $\sigma_0$ and $\sigma_f$ without knowing the if there are measurement between them.} 
    	   	  	   		\label{between_two}
    	   	  	   	\end{figure}

 To illustrate evolution with collapses consider in a frame $\mathcal{F}$  two hypersurfaces $\sigma_0$ and $\sigma_f$ before and after a collapse at a point $x$, see figure \ref{between_two}.
 The wavefunction
  $\psi_{\sigma_f}$ is found by evolving the wavefunction to a hyperplane of collapse, applying the collapse operator and normalising, then evolving to $\sigma_f$:
  \begin{equation}
  \label{collapse_evo}
 \psi_{\sigma_0} \to \psi_{\sigma_f}  = \frac{U_{\sigma_t}^{\sigma_f}\hat{L}_{\sigma_t}(x)U_{\sigma_0}^{\sigma_t} \psi_{\sigma_0}}{\norm{\hat{L}_{\sigma_t}(x)U_{\sigma_0}^{\sigma_t} \psi_{\sigma_0}}}.
  \end{equation}
  It is necessary that  \textit{all points of collapse between $\sigma_0$ and $\sigma_f$ are known in order to construct such a map between them}, as in general $\hat{L}_{\sigma_t}(x) \psi_\omega   \neq \psi_\omega $ for any $\omega$. Therefore in order to relate wavefunctions in different frames  on  their respective constant time hypersurfaces all collapses between those hypersurfaces must be known.
    	   	  	   	
   To find the condition on $\hat{L}_\sigma(x)$ for consistency with relativity we consider the probability $P (x_1, x_2 | \psi_{\sigma_{1}})$ which is  Eq. \eqref{covar_prob}  applied to  the case of two measurements at space-time points $x_1$ and $x_2$ given an  initial wavefunction $\psi_{\sigma_{1}}$\footnote{To keep notation simple and to highlight the invariance requirements we write  $P_{t_1} ( \mathbf{x}_1, | \psi_{\sigma_{1}})$ as $P (x_1, | \psi_{\sigma_{1}})$ however as $x_1$ is a space-time point of measurement the equation below should be understood in the same way Eq. \ref{covar_prob} is. } .  $\sigma_{1}$ is a constant time hypersurface intersecting the point $x_1$ in $\mathcal{F}$. For quantum mechanics with measurements the wavefunction $\psi_{\sigma_{1}}$ can be assumed to be specified by measurements in the past of $\sigma_1$. Then SR implies that:
  \begin{equation}
  \label{rela_requirement_2}
  P (x_1, x_2 | \psi_{\sigma_{1}}) = P (x'_1, x'_2 | \psi'_{\sigma'_{1'}}),
  \end{equation}

If the points $x_1$ and $x_2$ are time-like to each other and $x_1$ occurs before $x_2$ in all frames 
  the conditional probability for one frame is given by:
  \begin{equation}
  \label{expl_conditiona_prob}
  P (x_1, x_2 |\psi_{\sigma_{1}}) =  \norm{\hat{L}_{\sigma_2}(x_2)U_{\sigma_1}^{\sigma_2} {\hat{L}_{\sigma_1}(x_1)\psi_{\sigma_1}}}^2
  \end{equation}
  To compare the two sides of Eq. \eqref{rela_requirement_2} the relationship between the wavefunctions $\psi_{\sigma_{1}}$ and $ \psi'_{\sigma'_{1'}}$ must be specified. If there are no measurements (hence no collapses) between the two hypersurfaces then they can be related by:
   \begin{equation}
   \psi'_{\sigma'_{1'}}  =  \Lambda^\dagger U_{\sigma_1}^{\sigma'_{1'}} \psi_{\sigma_{1}}. \label{init_lorentz_time}
   \end{equation}
   
If there are measurements between $\psi_{\sigma_{1}}$ and $ \psi'_{\sigma'_{1'}}$ then the wavefunctions can be related with Eq. \eqref{collapse_evo} when measurement occurs and Eq. \eqref{init_lorentz_time} for subsequent evolution, using the appropriate positions and outcomes of measurements. In standard quantum mechanics this is acceptable as it includes the concept of observers performing measurements and recording the results. So all measurements between the two surfaces can be compared between two frames.
  Assuming that the Hamiltonian is covariant so that Eq. \eqref{unit_lorentz} holds then the right hand side of  Eq. \eqref{rela_requirement_2} can be written as:
  \begin{subequations}
  		\label{primed_cond_time}
  	\begin{align}
  	&P (x'_1, x'_2 | \psi'_{\sigma'_{1'}}) = \\ & \norm{\hat{L'}_{\sigma'_{2'}}(x_2')U_{\sigma'_{1'}}^{'\sigma'_{2'}} \hat{L'}_{\sigma'_{1'}}(x_1')\psi'_{\sigma'_{1'}}}^2=\\
  	&\norm{\hat{L'}_{\sigma'_{2'}}(x_2') \Lambda^\dagger U_{\sigma_2}^{\sigma'_{2'}}  U^{\sigma_2}_{\sigma_1}  U^{\sigma_1}_{\sigma'_{1'}} \Lambda \hat{L'}_{\sigma'_{1'}}(x_1') \Lambda^\dagger U_{\sigma_1}^{\sigma_1'}  \psi_{\sigma_1'} }^2,
  	\end{align}
  \end{subequations}
  where Eq. \eqref{unit_lorentz} has been used to transform the unitary operators and $\sigma'_{1'}$ and $\sigma'_{2'}$ are hypersurfaces of collapse intersecting $x_1$ and $x_2$ in frame $\mathcal{F'}$ and $x'$ is the same spacetime point in a  different coordinate system.
  So by inspection the condition for Eq. \eqref{rela_requirement_2} to hold is:
  \begin{equation}
  \hat{L'}_{\sigma'_{t'}}(x') = \Lambda^\dagger U_{\sigma_{t}}^{\sigma'_{t'}} \hat{L}_{\sigma_{t}}(x) U_{\sigma'_{t'}}^{\sigma_{t}}\Lambda. \label{collapse_lorentz}
  \end{equation} 
  Eq. \eqref{collapse_lorentz} requires that the collapse operator transforms covariantly and that the collapse can be described by a operator acting on any space-like hypersurface intersecting $x$. This is equivalent to requiring that the collapse happens instantaneously in all inertial frames. \\

  If, instead $x_1$ and $x_2$ are space-like to each other then in some frames their time ordering may be reversed. In this case, if in $\mathcal{F}'$ $x_1$ precedes $x_2$ then Eq.  \eqref{expl_conditiona_prob} holds and in the primed frame we have: 
    \begin{multline}
    	\label{primed_cond_space}
    	P (x'_1, x'_2 | \psi'_{\sigma'_{1'}}) = \\ \norm{\hat{L'}_{\sigma'_{1'}}(x_1')U_{\sigma'_{2'}}^{'\sigma'_{1'}} \hat{L'}_{\sigma'_{2'}}(x_2') U^{'\sigma'_{2'}}_{\sigma'_{1'}}\psi'_{\sigma'_{1'}}}^2.    \end{multline}
  Substituting in Eq. \eqref{init_lorentz_time} and Eq. \eqref{collapse_lorentz} it is found that for Eq. \eqref{rela_requirement_2} to be satisfied:
  \begin{equation}
  \label{mirco_causality}
  [\hat{L}_{\sigma_1}(x_1), U_{\sigma_2}^{\sigma_1}\hat{L}_{\sigma_2}(x_2)U_{\sigma_1}^{\sigma_2}]=0
  \end{equation}
  which is met if $\hat{L}_{\sigma}(x)$ satisfies the microcausality condition. 
  
  As discussed in section \ref{qm_and_sr} the wavefunction is a tool to calculate probabilities and in order to be consistent with special relativity the wavefunction must collapse instantaneously in every inertial frame. Therefore although we have written the collapse operator as acting on a constant time hypersurface in a particular frame, it could be written as a collapse operator acting on the Hilbert space of any space-like surface using the relationship:
\begin{equation}
    \label{trans_collapse}
    \hat{L}_{\sigma_t}(x) = U^{\sigma_t}_{\omega}\hat{Q}_{\omega}(x)U_{\sigma_t}^{\omega}
\end{equation}
where $\omega$ is any arbitrary space-like hypersurface intersecting the point $x$, and $\hat{Q}_{\omega}(x)$ is a collapse operator like $\hat{L}_{\omega}(x)$ that also satisfies Eq. \eqref{collapse_lorentz} and Eq. \eqref{mirco_causality}. 

  In order to check that Eq. \eqref{rela_requirement_2} is satisfied it has been implicitly assumed that in any one frame the time ordering between $x_1$ and $x_2$ is known. Otherwise it would not have been possible to write the explicit expressions of Eq. \eqref{expl_conditiona_prob}, \eqref{primed_cond_time}, and  \eqref{primed_cond_space}.

  As mentioned already,  standard quantum mechanics has the concept of observers comparing results, which  means that the order of measurements can be known between frames.
 If in one frame if observer A measured $x_1$ to be before $x_2$ and in another frame observer B measures the inverse then A and B can reconcile their conditional probability distributions and check consistency with SR.
  In this section we have found that condition for relativistic collapse using the conditional probability for two collapses, however it can be easily shown that this applies to any number of collapses.

   	\section{The original GRW model} \label{orgin_GRW}
   	We will now describe the original non-relativistic GRW model.
   	The original GRW model is a model for N distinguishable particles. For each particle the initial condition is the first point of  collapse $x_0$ and wavefunction $\psi_0$ at that time of collapse. The model then gives the probability distribution for the next point of collapse given the previous point. Hence the model is Markovian and each particle has a series of collapses.  For simplicity here we describe the model for a single particle, using a relativistic language to make clearer the similarities and differences between the original GRW model and its relativistic generalization.  
   	
	Collapses occur randomly in time and are a realization of a Poisson point process with mean time $\tau$. Let $\Delta T_i$ be the time interval between the $(i-1)$-th and the $i$-th collapse. Since time intervals are absolute in Galilean 	relativity, there is no need to specify with respect to which frame they are defined. This situation will change when we will consider  relativistic generalizations.  
	
	Consider a one particle wavefunction $\psi_{\sigma_0}$ defined on some initial hyperplane $\sigma_0$, where the first collapse has occurred; we are now specializing the description to a frame where $\sigma_0$ refers to time $t  = 0$.
   		  	\begin{figure}
   		  		\centering
   		  		\includegraphics[width=0.7\columnwidth]{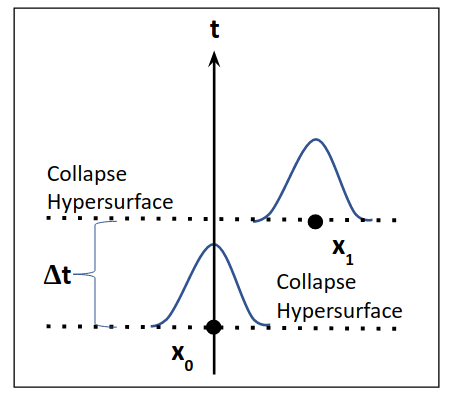}
   		  		\caption{Surfaces of collapse for the GRW model. The dotted black lines show the hypersurfaces where collapses occur. The solid lines show the amplitude of the wavefunction immediately after collapse. The initial collapse, $x_0$, is assumed to be at the origin, the next collapse will occur on a flat hypersurface after $\Delta t$.} 
   		  		\label{grw_tumulka}
   		  	\end{figure}
 
The next collapse will occur on the hypersurface $\sigma_{\Delta t}$, were $\Delta t = \Delta T_1$ as shown in figure \ref{grw_tumulka}. The probability distribution for a collapse to occur at a point $\textbf{x}_1 \in \mathbb{R}^3$ on this surface is:
   	\begin{equation}
   	P(\textbf{x}_1 | x_0, \Delta t, \psi_{\sigma_0}  ) =  \norm{\hat{L}_{\sigma_{\Delta t}}(\textbf{x}_1) U_{\sigma_0}^{\sigma_{\Delta t}} \psi_{\sigma_0}  }^2 .
   	\end{equation} 
   	   	where $x_0$ is the spacetime  coordinates of the first point of collapse, $U_{\sigma_0}^{\sigma_{\Delta t}} $ is the standard unitary evolution between time 0 and time $\Delta t$, generated by the Schr\"odinger equation, and:
   	   	\begin{equation} 
   	   	\label{grw_op_col}
   	   	\hat{L}_{\sigma_{x}}(\textbf{x}) \coloneqq \left( \frac{\alpha}{\pi}\right)^{\frac{3}{4}} \exp\left[-\frac{\alpha (\textbf{x}- \hat{\textbf{q}})^2}{2} \right]
   	   	\end{equation} 
   	   	where $\alpha$ is a free parameter of the model, $\sigma_x$ is the hyperplane intersecting the point $(\textbf{x},t)$, and $\hat{\textbf{q}}$ is the position operator of the particle. 
   	This distribution is normalised such that:
   	\begin{equation}
   	\int_{\mathbb{R}^3} d^3\underline{x}  \: P(\textbf{x}_1 | x_0, \Delta t,  \psi_{\sigma_0}  ) = 1.
   	\end{equation}
Immediately after a collapse the wavefunction is localised about the point of collapse, as given by Eq. \eqref{collapse_evo}, which in this case reads:
\begin{equation} \label{eq:gdrfg}
\psi_{\sigma_{\Delta t}} \rightarrow \psi_{\sigma_{\Delta t}}^{\text{\tiny (c)}} = \frac{\hat{L}_{\sigma_{\Delta t}}(\textbf{x}) U_{\sigma_0}^{\sigma_{\Delta t}} \psi_{\sigma_0}}{\| \hat{L}_{\sigma_{\Delta t}}(\textbf{x}) U_{\sigma_0}^{\sigma_{\Delta t}} \psi_{\sigma_0} \|}.
\end{equation}
Through this spontaneous collapse spatial superpositions are destroyed and hence classicality can emerge. In between collapses the wavefunction evolves according to the Scr\"{o}dinger equation. 

The above rules define {\it when}, {\it where} and {\it how} the collapses occur. We will use the same logic to define the relativistic generalization of the GRW model. 

\section{GRW and Special Relativity} \label{collapse_sr}  
The GRW model is a discrete time model which provides a conditional probability distribution for the position of a spontaneous collapse given the position of previous collapses. As the model is Markovian the conditional probability for a collapse only depends on the most recent collapse, not the whole prior series of collapses. We claim that any extension of GRW to the relativistic regime must provide a prescription for calculating the probability of the next point of collapse given the position of the previous point and that this probability distribution must be Lorentz invariant. Note that this definition of Markovianity assumes that there is a time ordering for the points of collapse.  

Additionally a relativistic GRW model must cause spatial superpositions of particles to collapse, and for $N > 1$ particles it must have an amplification mechanistic to ensure emergence of macroscopic classicality. 

We note that for a relativistic GRW model, an initial seed point of collapse must be given to define the model. This initial point breaks the Poincar\'e covariance, hence the appropriate symmetry group is the Lorentz group.

As already remarked on in section \ref{qm_and_sr},  for a spontaneous collapse model to be relativistic {\it both} the initial conditions between inertial frames must be comparable {\it and} the dynamics must be Lorentz covariant. The literature focussed in ensuring the second request to be satisfied, while here we will show that the first one in general is not, apart from specific situations.
 More specifically, in this section we will consider what these requirements imply for the form of a relativistic spontaneous collapse model for  a single particle, distinguishable particles and indistinguishable particles. 
 
\subsection{Relativistic condition for a single particle}

We consider a relativistic GRW model for a single particle. 
For a single particle there is a single series of collapses. In analogy with the original GRW model, one is tempted to define the times at which, in a given frame, collapses occur via a Poissonian distribution with average time $\tau$, but then one is faced with the fact that due to time dilation this prescription is not Lorentz invariant. In order to overcome this difficulty, the time intervals between collapses have to be defined in terms of Lorentz invariant time-like four-distances, this seems to be the only way to ensure that the time intervals are defined in a frame independent way. The four-distances have to be time-like not only because we are seeking a sequences of time intervals, but also because this prescription allows to define a time-ordered sequence of collapses. This is done for example in \cite{tumulka2006relativistic}. 

\begin{figure}
	   		\centering
	   		\includegraphics[width=1\columnwidth]{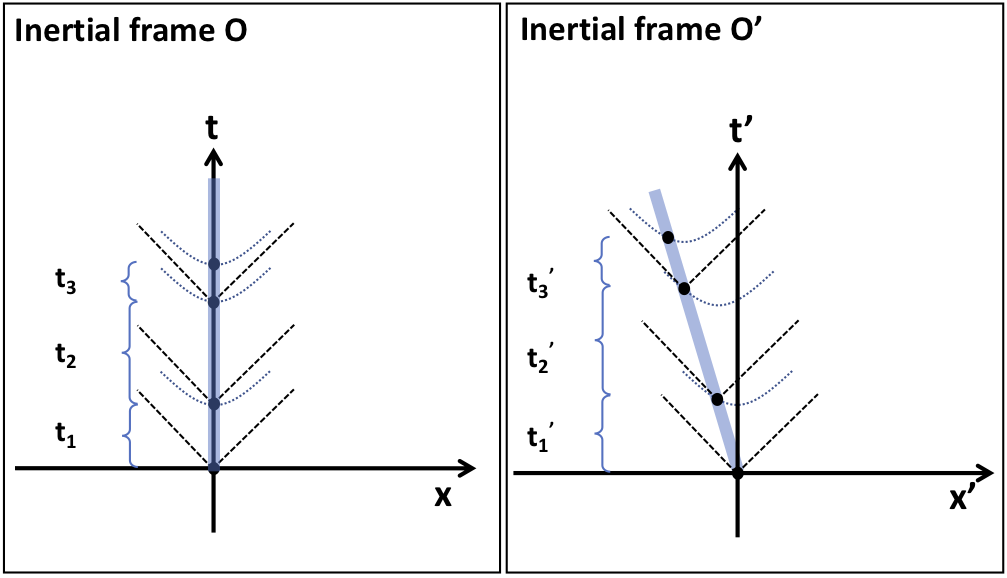}
	   		\caption{A schematic diagram showing how the stochastic process defines intervals between collapses. The shaded  area shows the maximum of the wavefunction's density, straight dotted lines show the future light cone of each point of collapse and the curved dotted lines show the surfaces of constant 4-distance from the previous point of collapse. The left diagram shows a frame where each collapse occurs at the same spatial point so the coordinate time and the 4-distance coincide, $\Delta T_i = t_i$. The right diagram shows a different inertial frame where the 4-distance between each point of collapse is still $ \Delta T_i$ but the coordinate time is different $t_i'$.   } 
	   		\label{time_inter}
	   	\end{figure}

Consider then a Poissonian point process with average $\tau$, with initial value $0$. Let $\Delta T_i$ be the distance between the $i$-th and $(i-1)$-th point of the process. Then define the times at which collapses occur as follows. Given the initial point of collapse  $x_0 = ({\bf x}_0, x^0_0)$, the next point of collapse $x_1 = ({\bf x}_1, x^0_1)$ will  occur at four-distance $\Delta T_1$ from $x_0$,  therefore $x_1$ will be on the future hyperboloid defined by all points with same {\it time-like} four-distance $\Delta T_1 = |x_1 - x_0|$ from $x_0$; the following point of collapse $x_2 = ({\bf x}_2, x^0_2)$ will lie  in the future hyperboloid defined by all points with same time-like four-distance $\Delta T_2 = |x_2 - x_1|$ from $x_1$, and so on. See Figure \ref{time_inter}.

The four-distances among consecutive collapses have an interesting physical interpretation.  Consider a particle whose wavefunction is well-localised in an inertial reference frame $O$ where the particle is at rest, for simplicity in the origin. In that frame, collapses are likely to occur only about the origin (where the wavefunction is non-zero), and the four-distances $\Delta T_i$   between consecutive collapses corresponds to the {\it coordinate} time intervals $\Delta t_i$ between collapses. In a different inertial frame $O'$, the particle will be moving and while the four-distances among the collapses do not change, the coordinate time intervals $\Delta t_i'$ are dilated. The opposite would be true for a well-localised particle at rest with respect to $O'$, thus in motion with respect to $O$. Therefore, the four-distances $\Delta T_i$ roughly correspond to the coordinate time intervals in the frame at rest with respect to the particle; in all other frames, the coordinate time intervals between collapses undergo time dilation.  So observers measure different rates of collapse in different frames due to time dilation, but the overall prescription of the rate of collapse is frame independent. This is analogous to the situation in particle physics where a particle with a half life $\lambda$, for example a muon, appears to have a longer half life when it is travelling at a high velocity inside a particle accelerator.

The prescription above defines, in a relativistic invariant way, {\it when}  a collapse occurs. The model must also define {\it where}  on the hyperboloid the collapse  occurs, i.e. give a normalized probability distribution for the position of the collapse on that hypersurface, such as done for example in \cite{tumulka2006relativistic}. In the spirit of GRW, this probability distribution must be equal to $P_\Sigma(x) = \| \hat{L}_\Sigma (x) \psi_\Sigma \|^2$ in order to avoid superluminal signaling, where $\hat{L}_\Sigma (x)$ is the collapse operator centred around the point of collapse $x$, defined on the hyperboloid $\Sigma$. 
We can leave $\hat{L}_\Sigma (x)$ unspecified, but it has to be chosen in such a way that it localises the wavefunction, is Lorentz covariant and that the probability is correctly normalized. However once it is specified it defines the collapse operator on all space-like hypersurfaces through Eq. \eqref{trans_collapse}.

\begin{figure}
	   		\centering
	   		\includegraphics[width=0.78\columnwidth]{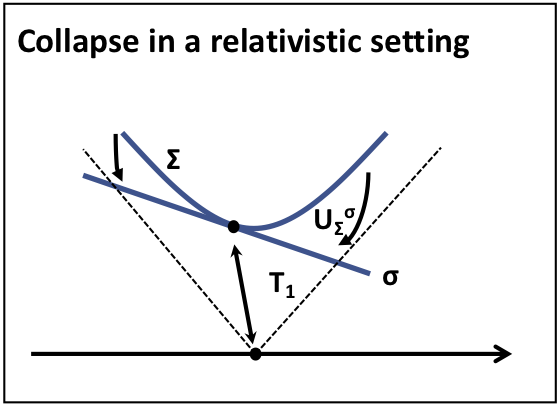}
	   		\caption{ Schematic diagram showing possible surfaces of collapse in relativistic GRW. The curved thick line labeled $\Sigma$ is a hyperboloid of made up of points  4-distance $\Delta T_1$ from the previous point of collapse. The collapse operator can be defined on the surface $\Sigma$ or equivalently on the hyperplane $\sigma$ via the operator  $U_\Sigma^\sigma$ The straight thick line labeled $\sigma$.}
	   		\label{tum_model_defn}
	   	\end{figure}

The last ingredient is {\it how} a collapse occurs, i.e. how the wavefunction changes due to a sudden collapse at $x$. \cite{tumulka2006relativistic} assumes that the wave function collapses along the hyperboloid previously introduced; this is mathematically implemented by applying $\hat{L}_\Sigma (x)$ to $\psi_\Sigma$, and then normalizing the collapsed wavefunction. In fact, the collapses can be carried out with respect to {\it any} space-like hypersurface containing the  point of collapse as the two prescriptions can be related by a unitary transformation. Specifically, suppose the collapse is defined to occur along a space-like hypersurface $\omega_1$ according to the prescription:
\begin{equation} \label{eq:gdrfg}
\psi_{\omega_1} \rightarrow \psi_{\omega_1}^{\text{\tiny (c)}} = \frac{ \hat{L}_{\omega_1}(x)\psi_{\omega_1}}{\| \hat{L}_{\omega_1}(x)\psi_{\omega_1} \|},
\end{equation}
where $x$ is the point of collapse. Given a second space-like hypersurface $\omega_2$ containing the point of collapse $x$, since $\psi_{\omega_2}  = U_{\omega_1}^{\omega_2} \psi_{\omega_1}$ for the wavefunction prior to the collapse, and $\psi_{\omega_2}^{\text{\tiny (c)}} = U_{\omega_1}^{\omega_2}\psi_{\omega_1}^{\text{\tiny (c)}}$ for the wavefunction after the collapse (because by construction there are no collapses in between $\omega_1$ and $\omega_2$ apart from $x$, since all collapses are assumed to be time-like separated with respect to each other), then Eq.~\eqref{eq:gdrfg} can be equivalently rewritten as:
\begin{equation} \label{eq:gdrfg2}
\psi_{\omega_2} \rightarrow \psi_{\omega_2}^{\text{\tiny (c)}} = \frac{\hat{L}_{\omega_2}(x)\psi_{\omega_2}}{\| \hat{L}_{\omega_2}(x)\psi_{\omega_2} \|},
\end{equation}
with $\hat{L}_{\omega_2}(x) = U_{\omega_1}^{\omega_2} \hat{L}_{\omega_1}(x) U_{\omega_2}^{\omega_1}$. See figure \ref{tum_model_defn}. Also the probability distribution $P_\Sigma(x) $ previously defined can be computed along any space-like hypersurface passing through $x$, since 
\begin{equation}
P_\Sigma(x) = \| \hat{L}_\Sigma (x) \psi_\Sigma \|^2 =  P_\omega(x) = \| \hat{L}_\omega (x) \psi_\omega \|^2 \equiv P(x),
\end{equation}
as one can easily check. It is in this sense that we can say that the collapse can be described consistently in all frames.

 Therefore we are precisely in the same situations envisaged by Albert and Aharonov: {\it a collapse occurs instantaneously  along all space-like hyper-surfaces intersecting the point of collapse}, with the only (important) difference that there the collapses are triggered by measurements, while here they are part of the dynamical law. As pointed out by Albert and Aharonov, this is necessary so that every inertial observer can provide a normalized wavefunction both before and after the collapse on their constant-time hyperplanes.  Constant-time hyperplanes are important because these are the hypersurfaces  where observers describe their physics.

It is for this reason that the model presented in \cite{dove1996local} is not a successful relativistic model, as this model has the wavefunction collapse only in the future light cone of the point of collapse. This means that the state is not normalised along different constant time hyperplanes and hence the theory does not give normalised probability distributions for systems with entangled particles. 

We argue that the only consistent way to understand the model in \cite{tumulka2006relativistic} is that the wavefunction collapses on every hypersurface intersecting the point of collapse, i.e. it collapses instantly in every frame. This is the only consistent way to interpret the model as otherwise the wavefunctions on hyperplanes after a point of collapse would be ill defined. This in agreement with Eq. 37 of \cite{tumulka2006relativistic} which gives the wavefunction on a constant time hypersurface for given foliation of spacetime.

To see how an inconsistency would arise otherwise consider a collapse at point $x$ and three hypersurfaces of interest, a hyperboloid  $\Sigma$ intersecting $x$, a hyperplane  $\sigma_{t_1}$ intersecting the point $x$ and  a hyperplane $\sigma_{t_2}$ a short time in the future of  $\sigma_{t_1}$.
The state $\psi_{\Sigma}$ will be the collapsed state. Suppose  that collapses only occur on hyperboloids then the state on  $\sigma_{t_1}$ would be uncollapsed.  Now the question is: what is the state on  $\sigma_{t_2}$? 
As there is unitary dynamics everywhere except on the hyperboloid then the state can either be written as $\psi_{\sigma_{t_2}} = U^{\sigma_{2}}_{\sigma_{1}} \psi_{\sigma_{t_1}}$ meaning $\psi_{\sigma_{t_2}}$ would be an uncollapsed state or as $\psi_{\sigma_{t_2}} = U^{\sigma_{2}}_{\Sigma} \psi_{\Sigma}$  meaning $\psi_{\sigma_{t_2}}$ would be collapsed. The only way to resolve this inconsistency and still have unitary dynamics is to have the collapse occur along $\sigma_{t_1}$ as well.

Now we are in the position to assess whether this framework for a relativistic GRW model is consistent with special relativity. Given the initial wavefunction $\psi_{\sigma_0}$ defined on a space-like hyperplane $\sigma_0$ and the initial point of collapse $x_0$ on $\sigma_0$, the probability for the next collapse to occur at $x$ is given by 
\begin{equation}
\label{prob_single_collapse}
P (x_1 | x_0, \psi_{\sigma_0}) = \norm{\hat{L}_{\omega}(x_1)U_{\sigma_0}^{\omega} \psi_{\sigma_0} }^2
\end{equation}
where $\omega$ is a surface intersecting $x_1$. 
This conditional probability distribution is analogous to Eq. \eqref{covar_prob}  where $\psi_{\sigma_0}$  gives all the possible information about the system at $(x_0,t_0)$ based on the position of previous collapses.

For a Lorentz transformed inertial frame $\mathcal{F}'$ with coordinates $x'$ the initial conditions are the point of last collapse $x_0'$  and the wavefunction on the hyperplane $\sigma'_{0'}$.
Therefore special relativity requires that:
\begin{equation}
\label{rela_requirement}
P (x_1 | x_0, \psi_{\sigma_0}) = P (x_1' | x_0', \psi'_{\sigma'_{0'}}).
\end{equation}
Here the condition of Eq. \eqref{covar_prob} is applied to spontaneous collapse models where the measurements replaced by points of  spontaneous collapse \footnote{We will keep the notation of section \ref{ts_with_measurements} with the understanding that now the time coordinate of $x$ is now probabilistic.}. For a single particle the wavefunction is specified by the point of last collapse $x_0$. 
As described in  section \ref{qm_and_sr},  in order to check this condition one must be able to {\it compare the initial conditions  } (here the wavefunction and position of the previous collapse) {\it between different inertial frames}, as noted in \cite{breuer1998relativistic}. 
This has consequences when considering collapse models for multiple particles, as we will see in sections \ref{rela_cond_dis} and \ref{rela_cond_indis}. 
	   	\begin{figure}
	   		\centering
	   		\includegraphics[width=1\columnwidth]{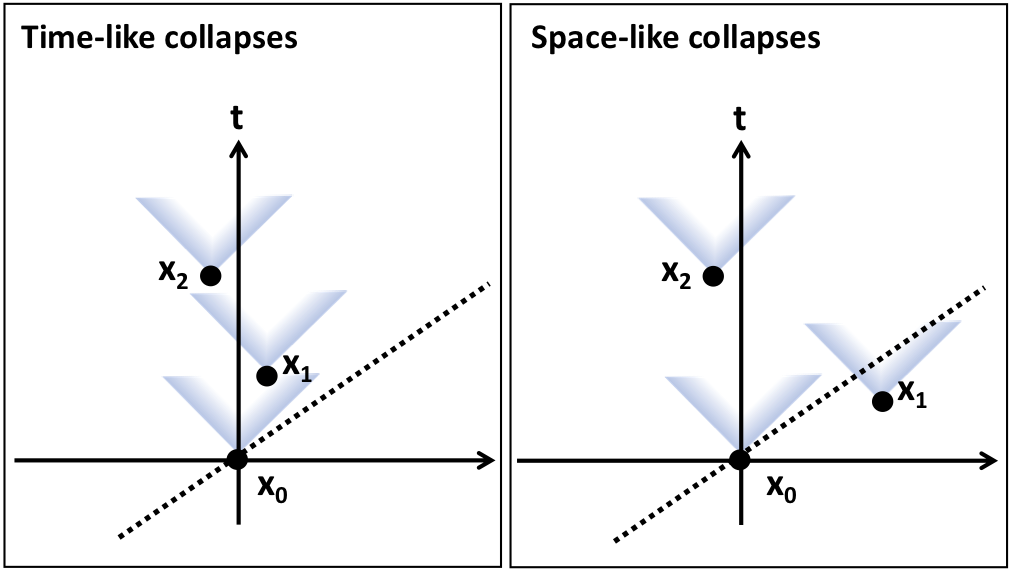}
	   		\caption{For time-like collapses (left) initial conditions between two inertial frames can always be related by unitary evolution as, for an initial collapse $x_1$, there can be no collapses between constant time hypersurfaces intersecting $x_1$, $\sigma_0$ and $\sigma'_{0'}$. For space-like collapses (right) then initial conditions between frames may not be related since there may be points of collapse between  $\sigma_0$ and $\sigma'_{0'}$  e.g. $x_1$.} 
	   		\label{space_like_collapse}
	   	\end{figure}

In order to verify Eq. \eqref{rela_requirement} the map between $\psi_{\sigma_0}$ defined on the constant-time hyperplane $\sigma_0$ for $O$,  and $\psi'_{\sigma'_{0'}}$ defined on the constant-time hyperplane $\sigma'_{0'}$ for $O'$ must be known, and in order to do this, positions of all collapses between those surfaces must be known.
For a series of {\it time-like} collapses this condition is met as there can be no collapses between $\sigma_0$ and $\sigma'_{0'}$, see figure \ref{space_like_collapse}, hence the two hyperplanes are related by Eq. \eqref{init_lorentz_time}. 
By the same argument presented in section \ref{ts_with_measurements}, the collapse operator $\hat{L}_{\sigma_t}(x)$ must transform as in Eq. \eqref{collapse_lorentz} and obey Eq. \eqref{mirco_causality}. 

 If these conditions are met these spatial collapses which are time-like to one another may be described in a way that is consistent with special relativity for a single particle.  The model proposed by \cite{tumulka2006relativistic} for a single particle meets these conditions.

On the contrary, for a single particle theory with collapses which are {\it space-like} to each other (we do not discuss how such a model could be formulated), the initial wavefunction in different inertial frames can no longer be related to each other by Eq. \eqref{init_lorentz_time}, as there might be collapses in the region enclosed between  $\sigma_0$ and $\sigma'_{0'}$, as shown in figure \ref{space_like_collapse}. Then to verify Eq. \eqref{rela_requirement} the position of all collapses in this region must be known, since this region includes points which are in the future of $x_0$ in $\mathcal{F}$; this is not possible.

One should notice the difference between standard quantum mechanics and spontaneous collapse models. Standard quantum mechanics has space-like collapses. However as discussed in section \ref{ts_with_measurements} this is consistent with relativity due to the position of collapses being given. In spontaneous collapse models the position of collapses are probabilistic and are not know \textit{a priori} and hence it cannot be taken for granted that initial conditions in two different inertial frames can be related. For space-like spontaneous collapses comparing initial conditions between two inertial frames is equivalent to requiring knowledge of future points of collapse in one of the inertial frames, as there might be collapses between two constant time hypersurfaces, see figure \ref{space_like_collapse}. Stochastic theories cannot meet this requirement, as the points of collapse are a single realisation of a random process and hence cannot be determined with certainty.

The convention that the initial collapse occurs at the origin has been taken. Since this is just a choice of coordinate system one would expect that the results discussed hold regardless of the choice of origin.  Since in two different inertial frames  $\mathcal{F}$ $(\mathcal{F}')$ the initial conditions are an initial point of collapse $x$ ($x'$) displaced from the origin and a wavefunction  $\psi_{0}$ ($\psi_{0'}$) on the hypersurface intersecting it, then the same rules for relating the two initial conditions as in the case of collapse at the origin can be applied. \\
So single particle spontaneous collapse models can meet condition Eq. \eqref{rela_requirement} when collapses are time-like to each other but for space-like collapses the initial condition for observers in two frames cannot be compared so the condition is not satisfied. For the model proposed in \cite{tumulka2006relativistic} for a single particle the collapses are time-like and hence it is a viable relativistic GRW model.  
Of course one does not expect space-like collapses for single particles as this would imply superluminal velocities, however this observation is relevant for multi-particle spontaneous collapse models. 
\subsection{Relativistic condition for N distinguishable particles} \label{rela_cond_dis}
The natural generalization of the previous model to the $N$  distinguishable particle case is to assume that there are $N$ series of collapses and hence $N$ realisations of the stochastic process. The $i^{th}$ realisation is:  $S_i = \{ T_{i1}, T_{i2}... \} $. For each realization, the construction of the collapse process---where they occur and how they change the wavefunction---is the same as for the single particle case. Note that in general, points of collapse associated to different particles can be space-like separated, while points of collapse associated to the same particle are always time-like to each other.
The Hilbert space for N distinguishable particles  is given by:
	\begin{equation}
	\label{multi_Hilbert}
	H = \underbrace{H_{1} \otimes H_2...\otimes H_{N}}_N,
	\end{equation}
	where $H_{i}$ is a single particle Hilbert space for the $i^{th}$ particle. The wavefunction on a hypersurface may be written as $\Psi_\sigma \in H $.

Then in this case,  the condition for consistency with special relativity is that:
\begin{multline}
P(x_{11},...., x_{i1}, ...x_{N1} | x_{10},...,x_{i0} ,...,x_{N0}, \Psi_{\sigma_0}) = \\
 P(x'_{11},...., x'_{i1}, ...x'_{N1} | x'_{10},...,x'_{i0} ,...,x'_{N0}, \Psi'_{\sigma'_{0'}})
 \label{n_prob}
\end{multline}
where for the $i^{th}$ series of collapses the collapse at $x_{i0}$ is followed by a collapse at $x_{i1}$ and $ \Psi_{\sigma_0}$ is the multiparticle wavefunction on a constant time hyperplane at the initial time. In the single particle case the initial wavefunction was defined on a hyperplane intersecting the initial point of collapse. As for multiple particles there are many initial points of collapse, it is not immediately obvious which hypersurface the initial wavefunction should be defined on.  The model can be defined consistently if in frame $\mathcal{O}$  the initial hypersurface intersects the earliest point of collapse in that frame for that generation of collapses, in this case the earliest point within the group $\{x_{10},...,x_{i0} ,...,x_{N0}\}$. So $\Psi_{\sigma_0}$ is the wavefunction on the hyperplane intersecting the earliest $x_{i0}$.  The relation Eq. \eqref{n_prob}  should hold true for every value of $j$, not only when $j=1$, however since the model is Markovian the relation can be  easily iterated and checked for any pair of consecutive collapses. 

A necessary requirement for Lorentz invariance of the probability distribution is that the distance between each $x_{ij}$  and $x_{ij+1}$  is a time-like 4 distance given by $T_{ij}$.

	As in the case of the single particle sector  in order for Eq. {\ref{n_prob}} to be satisfied then it must be possible to relate the initial wavefunctions $\Psi_{\sigma_0}$ and $\Psi'_{\sigma'_{0'}}$ in any two inertial frames. For a completely generic initial wavefunction $\Psi_{\sigma_0}$, which may be entangled, then collapses for any particle may affect the probability for the collapse of another particle via entanglement. 
	As collapses for different particles may be space-like to each other then the initial wavefunction in one frame $\Psi_{\sigma_t}$ and  cannot in general be related to $\Psi'_{\sigma'_{t'}}$, as is shown in section IV of \cite{aharonov1980states}.
	
		 	   	\begin{figure}
	 	   		\centering
	 	   		\includegraphics[width=0.8\columnwidth]{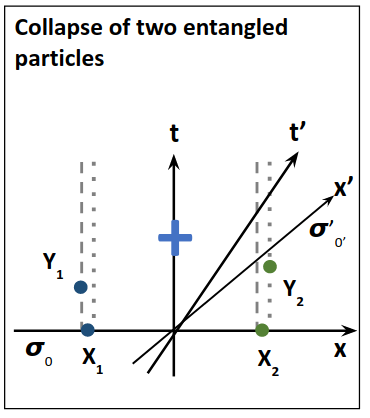}
	 	   		\caption{Schematic diagram showing two entangled distinguishable particles. The time and space axis are shown for two different frames. The dashed and dotted lines show the support for each part of the  wavefunction without collapses (see Eq.  \eqref{entangled_state}). Points $X_1$ and $X_2$ are the known initial points of collapse and $Y_1$ and $Y_2$ are possible future points of collapse.
 } 
	 	   		\label{entanged_no_compare}
	 	   	\end{figure}

	We will now give a simple illustrative example of an entangled initial state where two observers cannot compare initial conditions, but for a more rigorous explanation see \cite{aharonov1980states}. 
	Consider the situation shown in figure \ref{entanged_no_compare} with a system of two distinguishable particles. At time $t=0$ there are two collapses, at $X_1$ for particle 1 and at  $X_2$ for particle 2. We assume  by fiat that immediately after this the system is in the entangled state: 
	\begin{equation}
	\label{entangled_state}
	    |\psi_{\sigma_0} \rangle =  \frac{1}{\sqrt{2}}\Big( |L \rangle_1 |L \rangle_2 + |R \rangle_1 |R \rangle_2\Big)
	\end{equation}
	where the subscripts refer to the particle number and $| L \rangle $ is a localised state centred on the left and $| R \rangle $ is a localised state centred on the right. We assume that their centres are sufficiently far apart such that $\langle L |R \rangle \approx 0$. The entanglement ensures that any further collapse will localise both particles, for example if particle 2 collapses to $|R \rangle$ then particle 1 will also collapse to $|R \rangle$. This is a similar situation to the well known Bell locality scenario.
	
	For a GRW type model in frame $\mathcal{F}$ the probability of particle 1 collapsing at $Y_1$ can be given by knowing the state $ |\psi_{\sigma_0} \rangle$ and the position of the previous collapse $X_1$. However in frame  $\mathcal{F}'$ the initial state must be given on a constant time hypersurface ${\sigma'_{0'}}$ in that frame. As can be seen from figure \ref{entanged_no_compare} particle 2 may have already collapsed in $\mathcal{F}'$, for example at $Y_2$, which would also effect particle 1.
	
	So in order to compare the initial states on ${\sigma_0} $ and ${\sigma'_{0'}}$ the position of collapses between them (in this case $Y_2$) must be known, which would include collapse in the future for frame $\mathcal{F}$. Since in principle it is not possible to specify the position of future collapses for a stochastic theory, the two initial conditions cannot be compared.

	In conclusion there cannot be a special relativistic GRW model for entangled distinguishable particles. For the special case of non-interacting particles in a separable wavefunction it is possible to have a relativistic GRW. We now discuss the two cases more in detail.

	\subsubsection{Non-interacting separable particles}
For non-interacting particles the unitary evolution operator between two surfaces, $\omega_1$ $\omega_2$ may be written: 
	\begin{equation}
	\label{multitime_unitary} 
	W_{\omega_1}^{\omega_2} =   \underbrace{ U_{\omega_1,1}^{\omega_2} \otimes  U_{\omega_1,2 }^{\omega_2} \otimes U_{\omega_1,i }^{\omega_2}...}_{\text{N terms}} 
	\end{equation}
	Where $ U_{\omega_1,i}^{\omega_2}$ is the unitary operator for the $i^{th}$ particle. 
	The collapse operator for the the $i^{th}$ particle is:
		\begin{equation}
		\label{multi_particle_j_operator}
		\hat{L}_{\omega, i}(x) = \underbrace{\mathbb{I} \otimes \mathbb{I}...}_{\text{i terms}} \otimes \hat{L}_{\omega}(x) \otimes \underbrace{...\mathbb{I} \otimes \mathbb{I}}_{\text{N-i-1 terms}}
		\end{equation}
	where $\hat{L}_{\omega}(x)$ is the collapse operator for a single particle  (here for simplicity we assume that the form of each collapse operator is the same for every particle). There are N of such operators. 
		
		For a \textit{separable} initial condition:
			\begin{equation}
			\label{separable} 
			\Psi_{\sigma_0} =  \psi_{\sigma_0, 1} \otimes \psi_{\sigma_0, 2}...\otimes \psi_{\sigma_0, N}
			\end{equation}
		 then each side of  Eq. \ref{n_prob} can be factorised into $N$ distributions of the form:
\begin{equation}
\label{prob_dis_collapse}
P (x_{i1}, | x_{i0}, \psi_{\sigma_0, i} ) = \norm{\hat{L}_{\sigma_{i1}} (x_{i1})U_{\sigma_0, i}^{\sigma_{i1}}  \psi_{\sigma_0, i} }^2. 
\end{equation}
where $\sigma_{ij}$ is the hyperplane intersecting the point $x_{ij}$.
For consistency with special relativity	 each  $P (x_{i1}, | x_{i0}, \psi_{\sigma_0, i} ) $ must satisfy Eq. \eqref{rela_requirement}. 
If each particle has a series of collapses that are time-like to each other, and the collapse operator $\hat{L}_{\omega, i}(x)$ transforms as in Eq. \eqref{collapse_lorentz} then the model is consistent with special relativity. The model presented in \cite{tumulka2006relativistic} with a separable initial wavefunction meets this condition. 

If the initial wavefunction  $\Psi_{\sigma_0}$ is \textit{not separable} then Eq. \eqref{n_prob} will not be factorable. If it is not factorable then the initial wavefunction $	\Psi_{\sigma_1}$ cannot be specified in a frame independent way and hence the model cannot be consistent with special relativity.

\subsubsection{Interacting particles}
If the particles are interacting and  the wavefunction is initially separable then the unitary operator cannot be decomposed as in Eq. \eqref{multitime_unitary}. In this case the condition for Eq. \eqref{n_prob} to be factorable is:
\begin{equation}
\label{commute_inter_cond}
	[W_{\omega_1}^{\omega_2},\hat{L}_{\omega, i}(x)] = 0
	\implies [\hat{H},\hat{L}_{\omega, i}(x)] = 0
\end{equation} 
where  $\hat{H}$ is the Hamiltonian for the system (both the free and interacting parts). As is well known, if an operator commutes with the Hamiltonian then it corresponds to a globally conserved quantity. Therefore if the condition of Eq. \eqref{commute_inter_cond} holds then $\hat{L}_{\omega, i}(x)$  is a global operator and hence cannot be a local function of the fields. In this case, it has been shown that the dynamics does not result in a successful collapse model \cite{Priv_Comm_energy}. For example if the collapse operator is $\hat{H}$ then the collapse rate is proportional to the distance between the energy eigenvalues of the system, see Eq. 21 of \cite{adler2002environmental}. For systems in spatial superpositions but with degenerate energy eigenstates then the model would not predict any collapse. This would fail to solve the measurement problem as it would not lead to a reduction in the wavefunction for situations where we observe that wavefunction collapses. 

Therefore, since Eq.  \eqref{n_prob}  is not factorable, then one is faced with the problem as the non-separable wavefunction had, the initial wavefunction $\Psi_{\sigma_0}$ will be different in different frames due to the interaction. Hence it is not possible to have a special relativistic GRW model for interacting distinguishable particles.

\subsection{Relativistic condition for indistinguishable particles}\label{rela_cond_indis}
A relativistic GRW model for indistinguishable particles must only have a single collapse operator $\hat{L}_\sigma (x)$ which acts over every particle, to preserve the particle interchange symmetry or anti-symmetry for bosons and fermions respectively. Due to this, indistinguishable particles have the same relativistic condition as a single particle, namely that the stochastic process gives the 4-distance between points of collapse and Eq. \eqref{rela_requirement}, where $\psi_\sigma$ is an element of a N particle Fock space. From this the requirements Eq.\eqref{collapse_lorentz} and Eq.  \eqref{mirco_causality}  are derived.  If the collapses are time-like to each other then  Eq. \eqref{init_lorentz_time} can be used as the initial condition in one frame, the position of last collapse $x_0$ and the wavefunction $\psi_{\sigma_0}$ can be related to the initial condition in a different frame.

Conversely if the collapses are space-like to each other then Eq. \eqref{init_lorentz_time} does not hold  and such a model is not consistent with special relativity.

 \subsection{Emergence of macroscopicality for indistinguishable particles} 
\begin{figure}
   		\centering
   		\includegraphics[width=0.7\columnwidth]{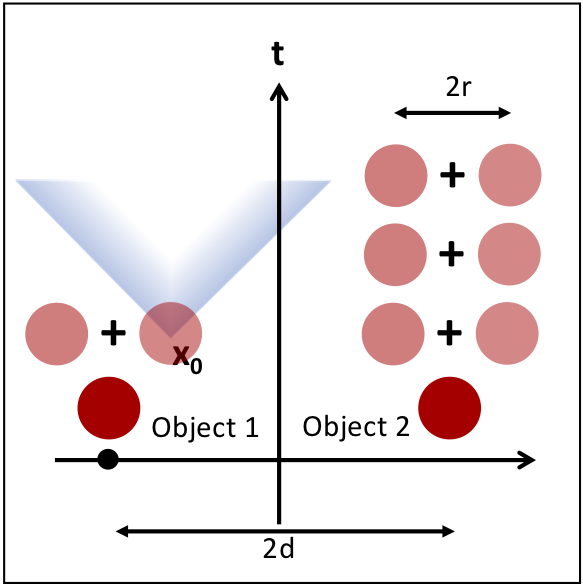}
   		\caption{Schematic spacetime diagram showing the evolution of a pair of space-like separated macroscopic objects separated by distance $2d$. For time-like collapses if there is a collapse at point $x_0$ the next collapse must occur in the future light cone of $x_0$ (shaded grey area), and therefore the object on the right will stay in a superposition.} 
   		\label{no_macro_blobs}
   	\end{figure}

 As has been discussed in section \ref{rela_cond_indis} a relativistic GRW model is possible for indistinguishable particles is possible if each collapse is time-like to the previous one. However such a model has an issue. If given a point of collapse $x_0$  the only region that the subsequent collapse can occur is in the future light cone of $x_0$ then macroscopic classicality is not recovered.
  	This can be seen with a simple example, with two macroscopic objects.
 Suppose there is a system made up a large number of indistinguishable particles N, where N is an  even number. The initial wavefunction of the system is two macroscopic objects, i.e. two areas with high densities of particles, with a large distance separation between the centre of mass of these two areas, labelled $2d$, see figure \ref{no_macro_blobs}. Assume that initially  each object is in a spatial superposition, separated by a distance $2r$, where $r \ll d$.  For simplicity we will work in one dimension but the argument can be extended to three dimensions. We will work in the framework of second quantization.

  The initial wavefunction of the system on a constant time hypersurface $\sigma_0$ is:
\begin{equation}
\label{intial_macro_super}
 |\Psi_{\sigma_0} \rangle = \frac{1}{2}	(\hat{A}_1 +\hat{A}_2)(\hat{B}_1+\hat{B}_2) |0 \rangle  
\end{equation} 
where $| 0\rangle$  is the vacuum of a N particle anti-symmetric Fock space and:
\begin{subequations}
	\begin{align}
	& \hat{A}_1 =\prod_{n=0}^{N/2} \hat{g} (-d-r,n) \\
	& \hat{A}_2 =\prod_{n=0}^{N/2} \hat{g} (-d+r,n) \\
	& \hat{B}_1 =\prod_{n=0}^{N/2} \hat{g} (d-r,n) \\
	& \hat{B}_2 =\prod_{n=0}^{N/2} \hat{g} (d+r,n) 
 	\end{align}
\end{subequations}
where: 
\begin{equation}
\hat{g}(x,n) = \hat{a}^\dagger(x-N \epsilon/4 +n \epsilon)
\end{equation}
	 	   	\begin{figure}
	 	   		\centering
	 	   		\includegraphics[width=0.9\columnwidth]{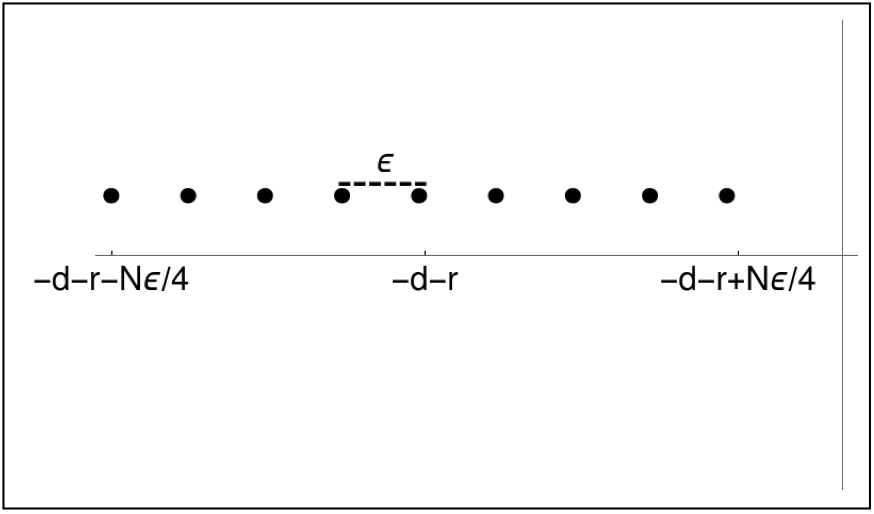}
	 	   		\caption{Diagram showing the action of the operator $\hat{A}_1$ on the vacuum, were particles are created separated by distance $\epsilon$.} 
	 	   		\label{a_1_op}
	 	   	\end{figure}
were $\epsilon$ is a distance such that $N \epsilon / 2 \ll r \ll d $. Additionally assume that the distance scale of the collapse is much less than the size of the superposition: $1/\alpha \ll r$. So the operator $\hat{A_1}$ acting on the vacuum creates $N/2$ fermions each displaced a distance $\epsilon$ from each other centred about the point $-d-r$ see figure \ref{a_1_op}, and similarly for $\hat{A_2}$, $\hat{B_1}$ and $\hat{B_2}$.

The number operator for the whole system is:
\begin{equation}
\label{number_total}
\hat{N}_T = \int_{-\infty}^{\infty}  dx  \:\hat{a}^\dagger(x) \hat{a}(x) .
\end{equation}
The number operator for the left part of the system is:
\begin{equation}
\label{number_left}
\hat{N}_A = \int_{-\infty}^{0}  dx  \:\hat{a}^\dagger(x) \hat{a}(x),
\end{equation}
With a the equivalent definition for $\hat{N}_B$. 
Finally the  number operator for the region to the left of $-d$ is:
\begin{equation}
\label{number_left_1}
\hat{N}_{A_1} = \int_{-\infty}^{-d}  dx  \:\hat{a}^\dagger(x) \hat{a}(x).
\end{equation}
The initial wavefunction $ |\Psi_{\sigma_0} \rangle$ is an eigenstate of the number operator for the total system:
\begin{subequations}
	\begin{align*}
	&\hat{N}_T  |\Psi_{\sigma_0} \rangle  \\
&=\int_{-\infty}^{\infty}  dx  \:\hat{a}^\dagger(x) \hat{a}(x)  \frac{1}{2}	(\hat{A}_1 +\hat{A}_2)(\hat{B}_1+\hat{B}_2) |0 \rangle  \\
&= \frac{1}{2}	(N \hat{A}_1 \hat{B}_1 +N \hat{A}_1 \hat{B}_2 + N \hat{A}_2 \hat{B}_1 + N \hat{A}_2 \hat{B}_2) |0 \rangle  \\
&= N  |\Psi_{\sigma_0} \rangle.
	\end{align*}
\end{subequations}
The initial wavefunction is also an eigenstate of the number operator for the left part of the system:
\begin{subequations}
\begin{align*}
	&\hat{N}_A  |\Psi_{\sigma_0} \rangle  \\
	&=\int_{-\infty}^{0}  dx  \:\hat{a}^\dagger(x) \hat{a}(x)  \frac{1}{2}	(\hat{A}_1 +\hat{A}_2)(\hat{B}_1+\hat{B}_2) |0 \rangle  \\
	&= \frac{N}{2} |\Psi_{\sigma_0} \rangle
\end{align*}
\end{subequations}
and similarly for the right part of the system $\hat{N}_B  |\Psi_{\sigma_0} \rangle = N/2 |\Psi_{\sigma_0} \rangle$. However the initial wavefunction is not in an eigenstate of the number operator for the region to the left of $-d$:
\begin{multline}
	\hat{N}_{A_1} |\Psi_{\sigma_0} \rangle  =\\
	\int_{-\infty}^{-d}  dx  \:\hat{a}^\dagger(x) \hat{a}(x)  	( \hat{A}_1 \hat{B}_1 + \hat{A}_1 \hat{B}_2 +  \hat{A}_2 \hat{B}_1 +  \hat{A}_2 \hat{B}_2) |0 \rangle  \\
	 	= \frac{1}{2}(\frac{N}{2} \hat{A}_1 \hat{B}_1 +  \frac{N}{2} \hat{A}_1 \hat{B}_2 +  \mathbb{I}  +  \mathbb{I} )|0 \rangle 
\end{multline}
which is not proportional to $|\Psi_{\sigma_0} \rangle$.
 $|\Psi_{\sigma_0} \rangle$ is also not an eigenstate of $\hat{N}_{A_2}$, $\hat{N}_{B_1}$ and $\hat{N}_{B_2}$. This implies there are two objects each in  a superposition over two areas, not one object in a superposition over four areas nor four objects each in a localised position. 
The amplification mechanism will cause a collapse of one of the objects almost immediately. Suppose  that the collapse is at spacetime point $(t, -d+r)$, where $t$ is so small that $U_{\sigma_0}^{\sigma_t} \approx \mathbb{I}$. Then following Eq. \eqref{collapse_evo} we find the wavefunction immediately after the collapse, on constant time hypersurface $\sigma_t$ to be:
\begin{equation}
\label{after_col}
| \Psi_{\sigma_t} \rangle = \frac{\hat{J}_{\sigma_t}(-d+r) | \Psi_{\sigma_0} \rangle}{\norm{\hat{J}_{\sigma_t}(-d+r) | \Psi_{\sigma_0} \rangle}^2}
\end{equation}
where $\hat{J}_{\sigma_t}(x)$ is an approximation for the form of a relativistic collapse operator $\hat{L}_{\sigma_t}(x)$ in the limit of low velocity particles. The form of $\hat{J}_{\sigma_t}(x)$ is:
\begin{multline}
\label{collapse_explicit}
\hat{J}_{\sigma_t}(x) | \Psi_{\sigma_0} \rangle = \\\int_{-\infty}^{\infty} dy  \: K(y) f_{\alpha}(x-y) \hat{a}^\dagger(y) \hat{a}(y) | \Psi_{\sigma_0} \rangle
\end{multline}
where $f_{\alpha}(x)$ is a function sharply peaked about $x=0$ with a width proportional to $1/\alpha$ and $K(y)$ is a normalisation function. This form ensures that particles are localised about the point of collapse. 
To evaluate Eq. \eqref{after_col} consider just the term:
\begin{subequations}
	\begin{align*}
 &\hat{J}_{\sigma_t}(-d+r)  \hat{A}_1 \hat{B}_1 | 0\rangle \\
 & = \frac{1}{2} \int_{-\infty}^{\infty} dy  \: K(y) f_{\alpha}(-d+r-y)  \times \\
 &\hat{a}^\dagger(y) \hat{a}(y)  \prod_{n=0}^{N/2} \prod_{m=0}^{N/2} \hat{g} (-d-r,n)  \hat{g} (d-r,m) | 0 \rangle
	\end{align*}
\end{subequations}
The contributions from the $\hat{g}(-d-r,n)$ and $\hat{g}(d-r,m)$ operators are weighted by factors of $f_\alpha(-2r+n\epsilon/4)$  and $f_\alpha(2d-2r+n\epsilon/4)$ respectively. As  $-2r+n\epsilon/4 \gg 1/ \alpha$ and $2d-2r+n\epsilon/4  \gg 1 / \alpha$ then $f_\alpha(-2r+n\epsilon/4) \approx 0$  and $f_\alpha(2d-2r+n\epsilon/4) \approx 0$. Hence :
\begin{equation}
\hat{J}_{\sigma_t}(-d+r) \hat{A}_1 \hat{B}_1 | 0\rangle \approx 0.
\end{equation}
A similar suppression occurs for $\hat{J}_{\sigma_t}(-d+r) \hat{A}_1 \hat{B}_2 | 0 \rangle $.
 However the terms $\hat{A}_2 \hat{B}_1 +\hat{A}_2 \hat{B}_2 $ are not suppressed as the $f_\alpha$ is approximately 1 for the part of the wavefunction centred on $-d+r$. Therefore we are left with:
 \begin{multline}
 \label{collapse_final_appox}
 \hat{J}_{\sigma_t}(-d+r) | \Psi_{\sigma_0} \rangle \approx \frac{N}{4}  \hat{A}_2 (\hat{B}_1 + \hat{B}_2) | 0 \rangle
 \end{multline}
therefore:
 \begin{equation}
 \label{after_collapse}
| \Psi_{\sigma_t} \rangle \approx  \frac{1}{\sqrt{2}} \hat{A}_2 (\hat{B}_1 + \hat{B}_2) | 0 \rangle
 \end{equation}

 So object 1 has been collapsed but object 2 remains in a superposition. Object 2 will be left in a superposition for  approximately $2d/c$ seconds, where $c$  is the speed of light, as can be seen in figure \ref{no_macro_blobs}. If $d$ is sufficiently large then one of the macroscopic objects will remain in a spatial superposition for an arbitrarily long time, in violation of what we observe in nature. 

To avoid this problem for macroscopic objects then a collapse model must permit space-like points of collapse. If there are space-like collapse points then the position of the initial collapse does not limit the region of possible collapses. Hence any region with a high-average number density of particles is almost certain to have a collapse occur within it in a short time interval. 
However as discussed in section \ref{rela_cond_indis}, if one attempts to include space-like collapses into the indistinguishable particle model suggested here then the model is not consistent with special relativity.

\section{Conclusion}
In this work we have considered the GRW model and its consistency with special relativity.  We have emphasised that for a model to be consistent  with special relativity  the dynamics must be Lorentz covariant and initial conditions in different inertial frames  must be able to be be related, we have applied these requirements for the case of relativistic GRW models. For a relativistic quantum theory this means that the initial wavefunction on a constant time hypersurface must be able to related between different frames. In table \ref{con_table} we have summarised the conclusions of this work, showing for which cases a relativistic GRW model is possible. 

 We have shown that a relativistic GRW is possible for single particles and non-interacting, non-entangled distinguishable particles as due to the fact that the collapses for each particle are time-like to each other, the initial conditions can be related in different inertial frames. However for entangled non-interacting distinguishable particles as entanglement implies that space-like collapses for one particle can effect the probability of collapse of another particle the initial conditions in different inertial frames cannot be related. For indistinguishable particles either the collapses are space-like hence not compatible with special relativity, or the collapses are time-like and the recovery of macroscopic classicality is not for granted and hence such a model is not a viable collapse model. For interacting particles as the interactions can entangle the particles, the initial conditions in two frames cannot be related, hence there is not a relativistic GRW model for interacting particles. \\
 The question is then if \textit{any} spontaneous collapse model can be relativistic, whether that model describes point like collapses such as in GRW or other models such as continuous spontaneous collapse. The two requirements that collapses for interacting particles must be time-like to each other to preserve frame independent probability distributions and that collapses must be space-like to ensure that macroscopic objects remain localised  seem to imply a contradiction and therefore that such a model is not possible. Recent work by Adler \cite{adler2018connecting} supports this idea. If collapse models are not consistent with special relativity the effect of violations of the Lorentz symmetry should be investigated in order to be confronted by experiment.
 
 One thing to note is that this work only considers the fixed particle sector, and a completely relativistic collapse model must also describe changes in particle number. This is something that should be considered  for further work in this subject. 

\begin{table}[h]
    \centering
\begin{tabular}{|p{3.5cm}||p{2cm}|p{2cm}|}
 \thickhline
 \multicolumn{3}{|c|}{Summary of Conclusions} \\
 \thickhline
 Particle Type & Separable State& Entangled State\\
 \hline
 Single   & Yes    & N/A \\
 \hline
 N distinguishable non-interacting&  Yes  & No \\
 \hline
 N indistinguishable non-interacting &No & No\\
 \hline
 Interacting    &No & No\\
 \hline
\end{tabular}
    \caption{A table showing the regimes where a relativistic GRW model is possible.}
    \label{con_table}
\end{table}

\begin{acknowledgments}
CJ thanks L. Asprea, J.L. Gaona Reyes and  G. Gasbarri for their comments and insights. CJ and AB thank R. Tumulka for his correspondence and comments. 
CJ and AB acknowledge financial support from the H2020 FET Project TEQ (grant
n. 766900). AB acknowledges the COST Action QTSpace (CA15220).
AB and CJ acknowledge financial support from INFN and the University of Trieste.
\end{acknowledgments}

\bibliographystyle{unsrt}
\bibliography{grw_third_version_angelo_edit3}

\end{document}